\documentclass[preprint,aps,amsmath,superscriptaddress,nofootinbib,tightenlines]{revtex4}
\usepackage{bm}
\usepackage{epsfig}
\usepackage{amsfonts}
\usepackage{bbm}
\def\nslash{n\!\!\!\slash}
\def\bnslash{\bar n\!\!\!\slash}

\def\OMIT#1{}

\newcommand{\nn}{\nonumber} 

\newcommand{\bn}{{\bar n}}
\newcommand{\bea}{\begin{eqnarray}}
\newcommand{\eea}{\end{eqnarray}}

\newcommand{\beq}{\begin{equation}}
\newcommand{\eeq}{\end{equation}}

\begin{document}

%%%%%%%%%%%%%%%%%%%%%%%%%%%%%%%%%%%%%%%%%%
%Define Title, Author, Address, Preprint#

%\preprint{\vbox{ \hbox{CMU-HEP-03-06}   \hbox{FERMILAB-Pub-03/069-T} }}

\title{Anomalous dimensions of the double parton fragmentation functions } 

\author{Sean Fleming\footnote{Electronic address: fleming@physics.arizona.edu}}
\affiliation{Department of Physics, 
         University of Arizona,
	Tucson, AZ 85721
	\vspace{0.2cm}}

\author{Adam K. Leibovich\footnote{Electronic address: akl2@pitt.edu}}
\affiliation{Pittsburgh Particle Physics Astrophysics and Cosmology Center (PITT PACC)\\ Department of Physics and Astronomy, 
         University of Pittsburgh,
	Pittsburgh, PA 15217
	\vspace{0.2cm}}
	
\author{Thomas Mehen\footnote{Electronic address: mehen@phy.duke.edu}}
\affiliation{Department of Physics, 
	Duke University, 
	Durham,  NC 27708
	\vspace{0.2cm}}

\author{Ira Z. Rothstein\footnote{Electronic address: izr@andrew.cmu.edu}}
\affiliation{Department of Physics, 
        Carnegie Mellon University,
	Pittsburgh, PA 15213
	\vspace{0.2cm}}

\date{\today\\ \vspace{1cm} }

%%%%%%%%%%%%%%%%%%%%%%%%%%%%%%%%%%%%%%%%%%
%Create the title page

\begin{abstract}
Double parton fragmentation is a process in which a pair of partons produced in the 
short-distance process hadronize into the final state hadron. This process is important for quarkonium production when the transverse momentum is much greater than the quark mass.  Resummation of logarithms of the ratio of these two scales requires the evolution equations for double parton fragmentation functions (DPFF). In this paper we complete the one-loop evaluation of the anomalous dimensions for the DPFF.  We also
consider possible mixing between the DPFF and single parton power suppressed gluon fragmentation
and show that such effects are sub-leading.
\end{abstract}

\maketitle

\newpage

%%%%%%%%%%%%%%%%%%%%%%%%%%%%%%%%%%%%%%%%%%
%Main body of the paper
%%%%%%%%%%%%%%%%%%%%%%%%%%%%%%%%%%%%%%%%%%

%\section{Introduction}

The production of $J/\psi$, $\Upsilon$, and other quarkonium states in hadron colliders is an important test of our understanding of perturbative QCD. Many early calculations used a color-singlet model (CSM)  in which the heavy quark-antiquark pair ($Q\bar{Q}$) were assumed to be produced with quantum numbers identical to the final state hadron, e.g., $^3S_1^{(1)}$ for the $J/\psi$ and $\Upsilon$.  The CSM calculates the production cross section for producing a $Q\bar{Q}$ pair with the right quantum numbers, multiplied by the quarkonium wavefunction. 
This model had to be rejected when measurements of prompt $J/\psi$ production at large transverse momentum at CDF revealed order of magnitude discrepancies between the CSM and experimental data. The modern theory of quarkonium production is based on Non-Relativistic QCD (NRQCD)~\cite{Bodwin:1994jh}. In this theory, the $Q\bar{Q}$ pair can be produced in a state with arbitrary quantum numbers. The nonperturbative transition of the heavy quark-antiquark pair produced in the short-distance process into the final state is governed by the QCD multipole expansion, and hence is an expansion in $v$, where $v$ is the relative velocity of the heavy quark and antiquark. This theory allows for new color-octet production mechanisms that can naturally explain  the size of the cross section at large $p_\perp$~\cite{Braaten:1994vv} but fails to explain some features of the data, most notably the polarization of the $J/\psi$ at large $p_\perp$~\cite{CW}. For an up-to-date next-to-leading order analysis of quarkonium production with comparison to a wide variety of experimental results, 
see Refs.~\cite{Butenschoen:2012qr,Butenschoen:2012qh,Butenschoen:2012px}.

At large $p_\perp$ the dominant mechanism for quarkonium production is fragmentation, in which
we consider the probability to produce an ``on-shell'' parton which then has some probability to
hadronize to the quarkonium of interest.
To make sensible predictions for such processes one first needs to prove factorization, which, at least in this context of this paper,  implies that at large $p_\perp$ the cross section can be written as
the product of probabilities. The parton to hadron transition is encapsulated by a fragmentation function which obeys a DGLAP evolution equation that can be used to resum large logarithms of $p_\perp/m_Q$ that
arise in the cross section.
Within the framework of NRQCD, this resummation has been performed for single parton fragmentation contributions to quarkonium production~ \cite{Braaten:1993mp,Braaten:1993rw,Braaten:1994xb}. 

There are many power suppressed corrections to this result, which naively one might think
are numerically irrelevant. However, for quarkonium fragmentation there is a  particular power correction,
coming from double parton fragmentation, whose contribution can be large for moderate values of $p_\perp$.
While these corrections are down by $ m_Q^2/p_\perp^2$ they are enhanced by a factor of
$v^4$ compared to the single gluon fragmentation function. In fact, the double parton fragmentation
operators are the unique power correction with this enhancement.

 For the case of quarkonium production, logarithms of $p_\perp/m_Q$ in double parton fragmentation have not been resummed.
The double parton fragmentation function (DPFF) was first introduced in Refs.~\cite{KQS,Kang:2011zza}, where it was evaluated in perturbation theory. It was shown to account for 80\% of the next-to-leading order CSM cross section and furthermore yield longitudinally polarized $J/\psi$. In Ref.~\cite{Fleming:2012wy} we derived factorization theorems for the DPFF using Soft Collinear Effective Theory (SCET)~\cite{Bauer:2000ew,Bauer:2000yr} and computed the evolution equations for the DPFF. In that paper, we focused on color-singlet and color-octet $^3S_1$ production only. However, production in other spectroscopic channels such as $^1S_0^{(8)}$ and $^3P_J^{(8)}$ are also important to collider production of quarkonia, at least at moderate transverse momentum \cite{CL1,CL2}.  Furthermore, the DPFF can mix into power suppressed single parton
fragmentation functions. The purpose of this paper is to complete the leading order calculation of the anomalous dimensions for the DPFF.
%\section{DPFF definition}

The DPFF, $D_{i,a}^{Q\bar Q}(u,v,z)$, was  defined in Ref.~\cite{Fleming:2012wy} in terms of the following SCET matrix element:
\bea
\label{QQfragfun}
& & \langle 0| 
\bar \chi_{n^\prime, \omega'_2}   \Gamma^{(\nu)}_i \{ \mathbbm{1},T^A\} \chi_{n^\prime,  \omega'_1}
 {\cal P}^H_{n',\bn'\cdot p}
\bar \chi_{n^\prime,  \omega'_4}   \Gamma_{i (\nu) } \{ \mathbbm{1},T^A\}  \chi_{n^\prime,  \omega'_3} |0\rangle
\\
& & =  8\, \delta(\omega'_1-\omega'_2+\omega'_3-\omega'_4) \int \frac{dz}{z} \, du \, dv \,\delta(z-\frac{\bn'\!\cdot\! p}{\omega'_1-\omega'_2})
 \delta(v-1-z\frac{\omega'_2}{ \bn'\!\cdot\! p}) \delta(u-z\frac{\omega'_4}{ \bn'\!\cdot\! p})\nn\\
& &\hspace{30ex}\times z D_{i,a}^{Q\bar Q}(u,v,z) \nn \,,
\eea
where $a = 1,8$ indicates the color-structure of the operator, and $\Gamma^{(\nu)}_i \in \frac{1}{2}\{\bnslash',\bnslash' \gamma_5, \bnslash' \gamma^\nu_{\perp'}\}$, where $\gamma^\nu_{\perp'}= \gamma^\nu -n'^\nu \bnslash'/2 - \bn'^\nu \nslash'/2$.
The notation used here is the same as in Ref.~\cite{Fleming:2012wy} and we refer the reader to that paper for details.
The matching of DPFF operators onto NRQCD production operators is also discussed in Ref.~\cite{Fleming:2012wy}. At lowest order in $v$, the operators with Dirac structures $\Gamma_1$, $\Gamma_2$, and $\Gamma_3^\nu$ match onto
$^3S_1$, $^1S_0$, and $^3S_1$ NRQCD production operators, respectively. $\Gamma_1$ corresponds to longitudinal polarization of the
$Q\bar{Q}$ pair, while $\Gamma_3^\nu$ corresponds to transverse polarizations. $P$-wave operators can appear at higher order in the $v$ expansion. In Ref.~\cite{Fleming:2012wy}, only the Dirac structure $\Gamma_3^\nu =\frac{1}{2}\bnslash' \gamma^\nu_{\perp'}$ was considered. It does not mix with the other Dirac structures under renormalization. We will see below that $\Gamma_1$ and $\Gamma_2$ mix among themselves. The variables $z$, $u$, and $v$ are also defined in Ref.~\cite{Fleming:2012wy}:
\bea
\label{lcmfcv}
z &=& \frac{\bn'\!\cdot\! p}{\omega'_1-\omega'_2}=\frac{\bn'\!\cdot\! p}{\omega'_4-\omega'_3}\\
v &=& z\frac{\omega'_1}{ \bn'\!\cdot\! p}=1+z\frac{\omega'_2}{ \bn'\!\cdot\! p}\nn\\
u &=& z\frac{\omega'_4}{ \bn'\!\cdot\! p}=1+z\frac{\omega'_3}{ \bn'\!\cdot\! p} \,.\nn
\eea
Here,  $z$ corresponds to the hadron $H$'s fraction of the $Q\bar Q$ pair  light-cone momentum, and  $u$ and $v$ correspond to the fraction of the total $Q\bar Q$ light-cone momentum carried by each of the heavy quarks in the $Q\bar Q$ pair. Eq.~(\ref{QQfragfun}) can be inverted to obtain
\bea
\label{invQQfragfun}
D_{i,a}^{Q\bar Q}(u,v,z)&=& \frac{1}{8} \int d\omega'_1d\omega'_2d\omega'_3 d\omega'_4
\delta(\omega'_1-\omega'_2-\frac{ \bn'\!\cdot\! p}{z})\delta(\omega'_2-\frac{ \bn'\!\cdot\! p}{z}(v-1))\,\delta(\omega'_4-\frac{ \bn'\!\cdot\! p}{z}u)\nn \\
&&\times
 \, \langle 0| 
\bar \chi_{n^\prime, \omega'_2}   \Gamma^{i (\nu) } \{ \mathbbm{1},T^A\} \chi_{n^\prime,  \omega'_1}
{\cal P}^H_{n',\bn'\cdot p}
\bar \chi_{n^\prime,  \omega'_4}   \Gamma^i_{ (\nu) } \{ \mathbbm{1},T^A\}  \chi_{n^\prime,  \omega'_3} |0\rangle \,.
\eea

%\section{Anomalous Dimensions}

\begin{figure}
\begin{center}
\includegraphics[width=1.5in]{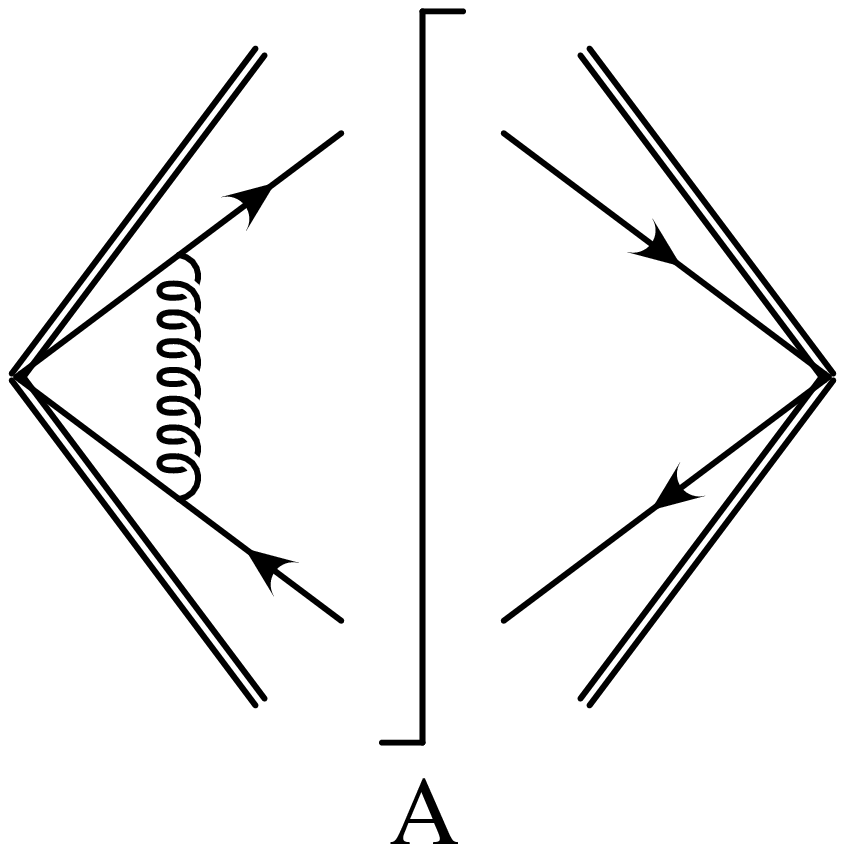}\quad\includegraphics[width=1.5in]{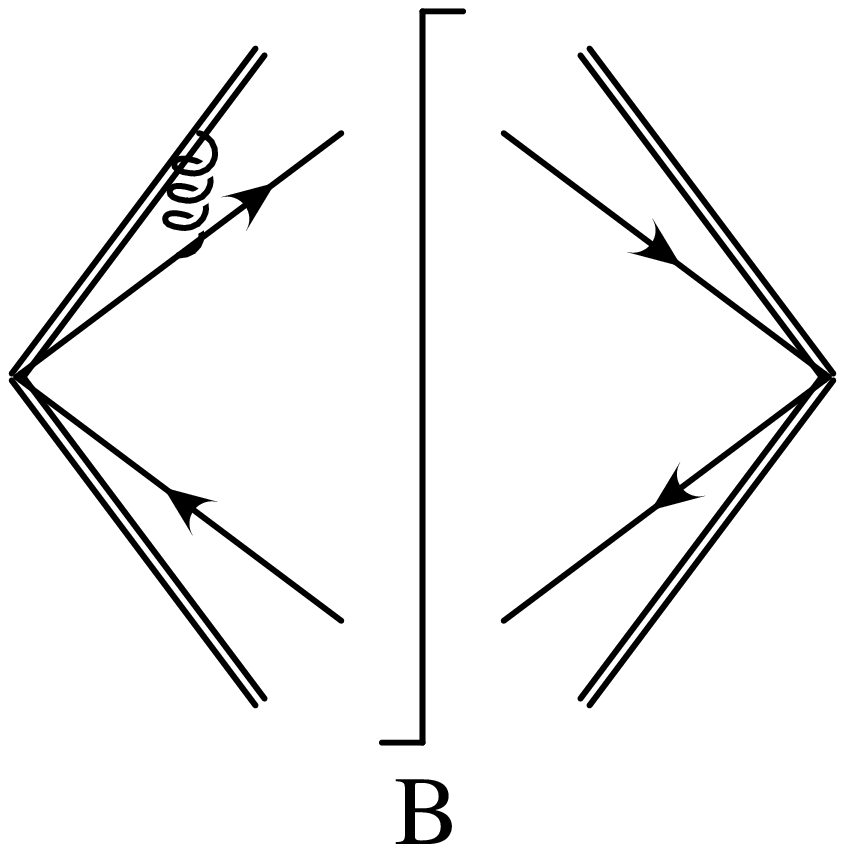}\quad\includegraphics[width=1.5in]{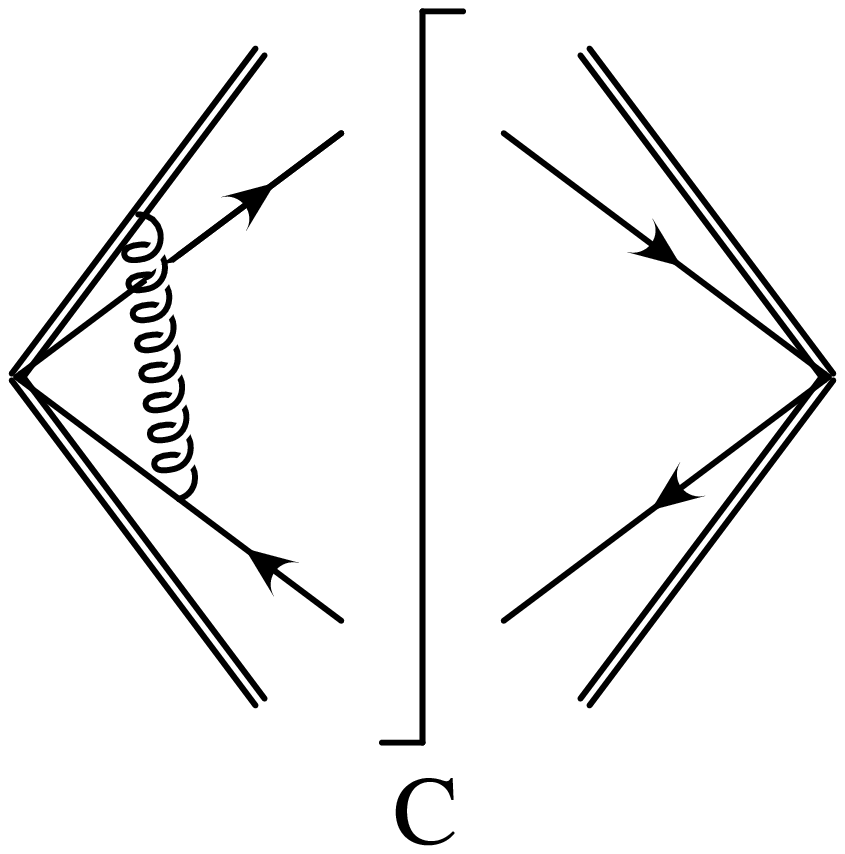}\quad\includegraphics[width=1.5in]{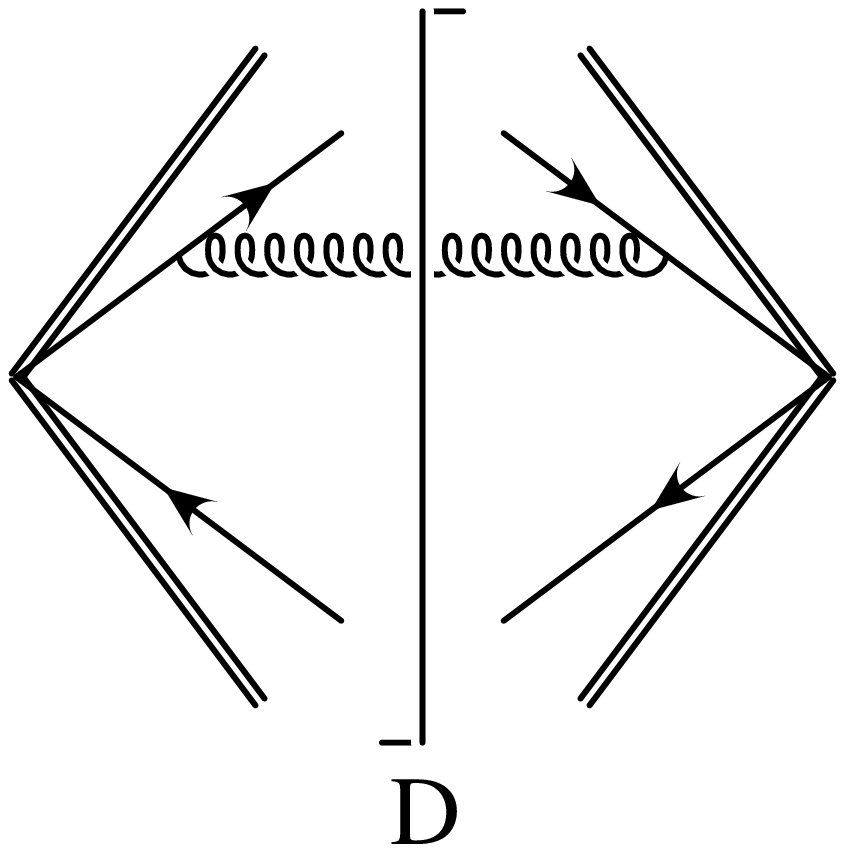}\\
\includegraphics[width=1.5in]{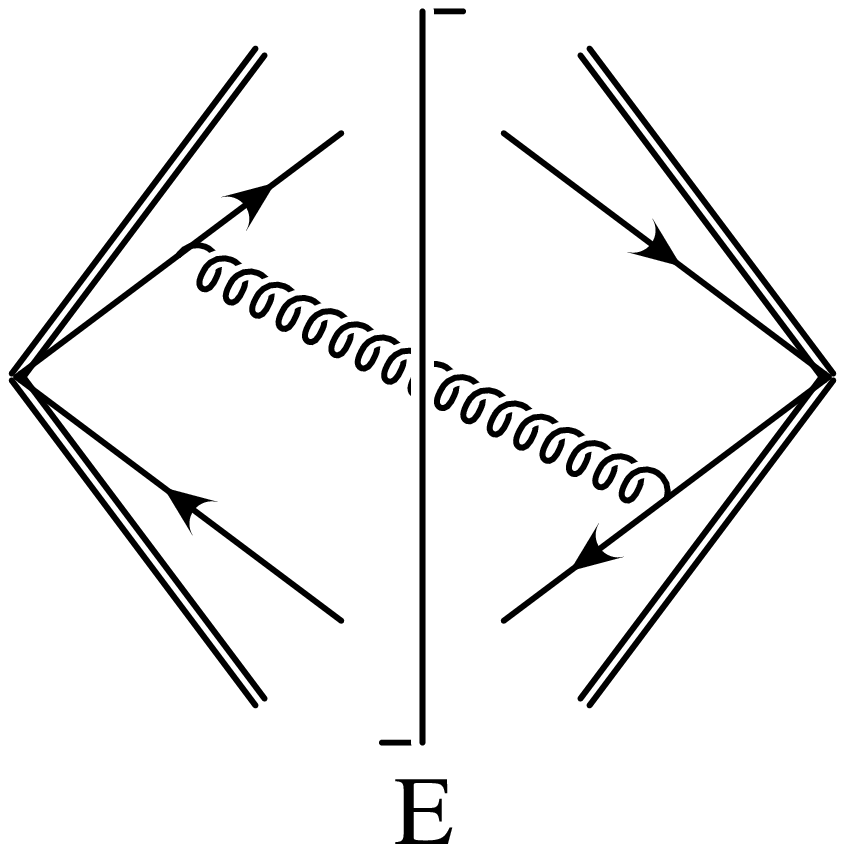}\quad\includegraphics[width=1.5in]{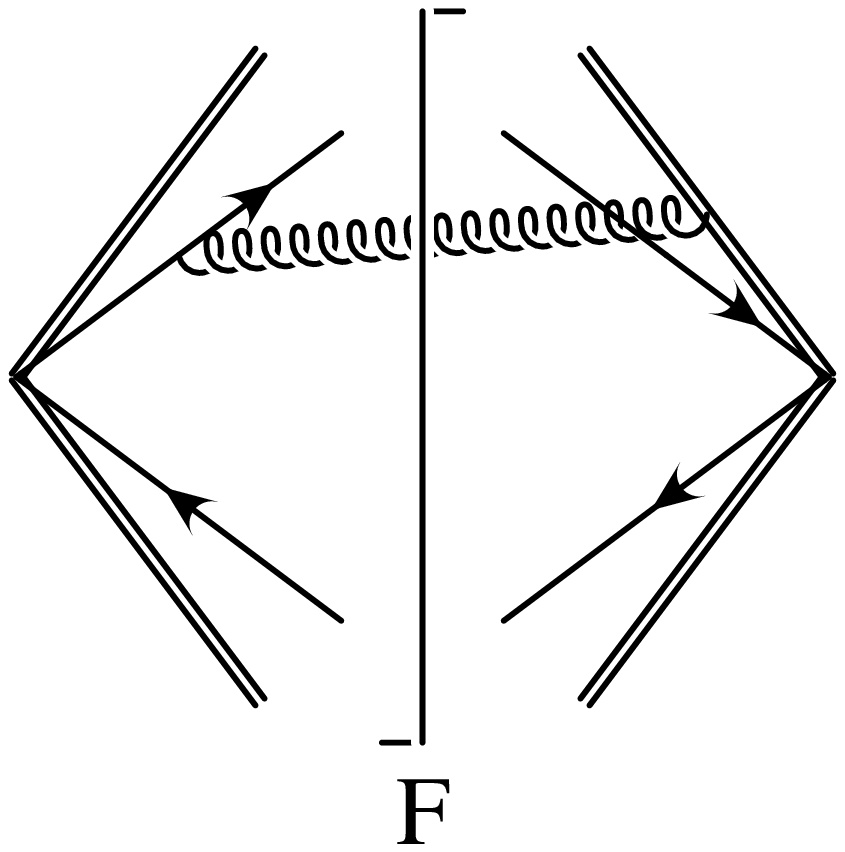}\quad\includegraphics[width=1.5in]{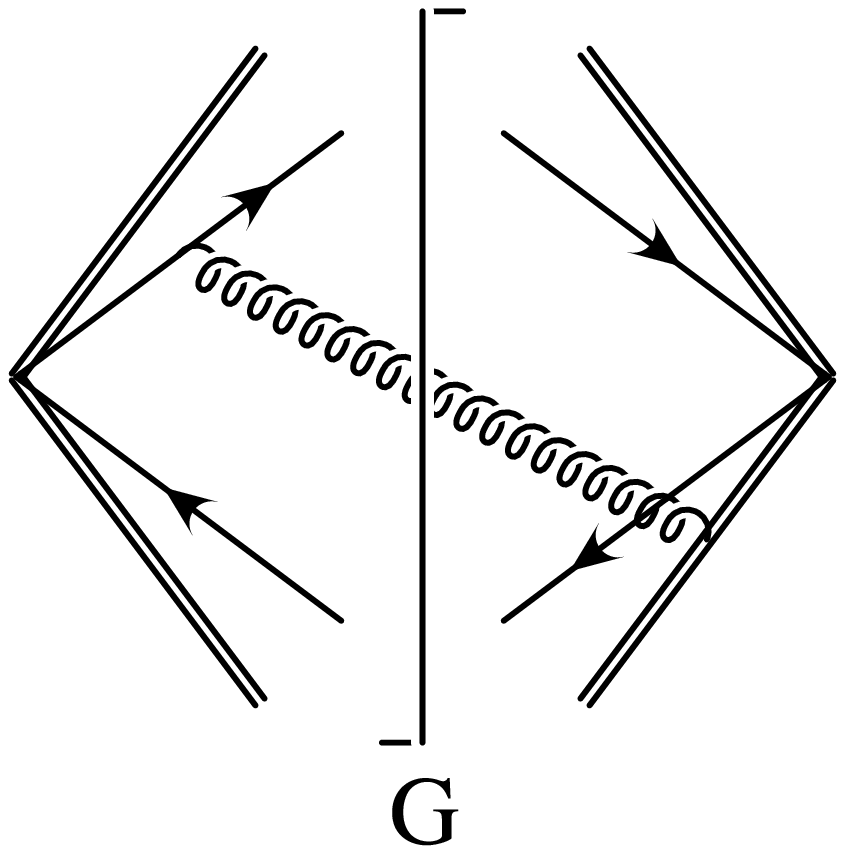}
\caption{The  one-loop diagrams required for computing the evolution of the DPFF. Not shown are the diagrams which are mirror images with respect to the horizontal and/or vertical axes.   }
\label{oneloopfd}
\end{center}
\end{figure}

The anomalous dimensions of the DPFFs are
determined from the ultraviolet (UV) counterterms to $D^{Q\bar Q}_{i,a}(u,v,z)$, which are calculated from the one-loop diagrams given 
in Fig.~\ref{oneloopfd}.
We work in Feynman gauge and use dimensional regularization to regulate UV divergences in loop diagrams. 
Here as in Ref.~\cite{Fleming:2012wy}, SCET collinear fields are represented by single lines and 
the double lines correspond to Wilson lines. There are additional diagrams that can be obtained by reflecting diagrams
in Fig.~\ref{oneloopfd} about the horizontal or vertical axis, or both axes. We refer to diagrams obtained  by reflecting about the horizontal axis by adding a hat, e.g. $\hat C$, diagrams obtained by reflecting about the vertical axis by adding a bar, e.g. $\bar C$, and by doing both reflections by adding a hat and bar, e.g. $\bar {\hat C}$. Note that $A = \hat A$, $D = \bar D$, and $\bar E =\hat E$ so these are not distinct diagrams. All other possible reflections lead to new diagrams.

In Fig.~\ref{oneloopfd}, we define the outgoing momentum of the quark and anti-quark on the left-hand side of the cut to be $p^\mu_4$ and $p^\mu_3$, respectively, and we define the incoming momentum of  the quark and anti-quark on the right-hand side of the cut to be  $p^\mu_1$ and $p^\mu_2$, respectively. The large components of these momenta are expressed in terms of the momentum fractions defined in Ref.~\cite{Fleming:2012wy}:
\bea
x &=& \frac{P}{\bn'\cdot (p_1+p_2)} \\
\lambda  &=& x \frac{\bn'\cdot p_4}{P} \nn \\
\xi &=& x \frac{\bn' \cdot p_1}{P} \, . \nn
\eea
Here $P$ is the large light-cone momentum component of the final state $Q\bar{Q}$ pair.

Diagram $A$ and its reflections  have different color factors depending on whether or not
the operator is color-singlet or color-octet. These color factors 
are  $\beta^{(1,8)}$, where $\beta^{(1)}=C_F$ and $\beta^{(8)}=-\frac{1}{2N_c} $. 
Evaluating diagram $A$ yields
\bea
M_A^{i,a} =  \frac{\alpha_s}{2\pi} \frac{\beta^{(a)}}{\epsilon} \bigg(\frac{z}{2P}\bigg)^3 \delta(1-z/x)\delta(\xi-v) O^{i,a} \left[\frac{u}{\lambda} \theta(\lambda-u) + \frac{\bar u}{\bar \lambda}\theta(u-\lambda)\right] \, .
\eea
On the right hand side the index  $a$ is not summed over.   We define $\bar u = 1-u$, $\bar v = 1-v$, $\bar \lambda = 1-\lambda$, and $\bar \xi = 1- \xi$, and the operator $O^{i,a}$ is  
\beq
O^{i,a} = \bar \xi_{n'} \Gamma_{i}\{ \mathbbm{1},T^A\} \xi_{n'}  \bar \xi_{n'} \Gamma_{i}\{ \mathbbm{1},T^A\} \xi_{n'}\, .
\eeq
The diagram related to $A$ by reflection about the vertical axis is obtained by making the replacements
\bea
\label{subs1}
M_{\bar A}^{i,a} = M_A^{i,a}(u\leftrightarrow v, \lambda \leftrightarrow \xi) \, .
\eea

The result of evaluating diagram $B$ is  
\bea
M_B^{i,a}&=& \frac{\alpha_s}{2\pi} \frac{C_F}{\epsilon}\bigg[ \frac{1}{\eta}+\ln\bigg(\frac{z \nu}{u P}\bigg)+1\bigg] \bigg(\frac{z}{2P}\bigg)^3 \delta(1-z/x)\delta(\lambda-u)\delta(\xi-v) O^{i,a} \, .
\eea
Note that this diagram exhibits a rapidity divergence that has been regulated using the regulator of Refs.~\cite{Chiu:2011qc,Chiu:2012ir}.
This is the origin of the factor of $1/\eta$, where $\eta$ is the rapidity regulator, as well as the scale $\nu$. We will see below that the rapidity divergences cancel in the sum over all diagrams. The diagrams related to $B$ by symmetry are obtained by making the following replacements:
\bea\label{subs2}
M_{\bar B}^{i,a} &=& M_B^{i,a}(u\leftrightarrow v, \lambda \leftrightarrow \xi), \\
M_{\hat B}^{i,a} &=& M_B^{i,a}(u\leftrightarrow \bar u, \lambda \leftrightarrow \bar\lambda), \nn \\
M_{\bar{\hat B}}^{i,a} &=& M_B^{i,a}(u\leftrightarrow \bar v, \lambda \leftrightarrow \bar\xi). \nn
\eea
Both color structures give the same result so the sum of all diagrams is
\bea
M_B^{i,a}+M_{\bar B}^{i,a}+M_{\hat B}^{i,a}+M_{\bar{\hat B}}^{i,a} &=& \\
&& \hspace{-12ex} 
\frac{\alpha_s}{2\pi} \frac{C_F}{\epsilon}
\bigg[ \frac{4}{\eta}+\ln\bigg(\frac{z^4 \nu^4}{u\bar u v \bar v P^4}\bigg)+4\bigg] \bigg(\frac{z}{2P}\bigg)^3 \delta(1-z/x)\delta(\lambda-u)\delta(\xi-v) O^{i,a}  \, .\nn
\eea

Diagram $C$ evaluates to
\bea
M_{C}^{i,a}&=& \\
&&-\frac{\alpha_s}{2\pi} \frac{\beta^{(a)}}{\epsilon}\bigg\{\bigg[ \frac{1}{\eta}+\ln\bigg(\frac{z\nu}{\bar u P}\bigg)\bigg]\delta(\lambda-u)-\frac{\bar u}{\bar \lambda}\frac{\theta(u-\lambda)}{(u-\lambda)_+}\bigg\} \bigg(\frac{z}{2P}\bigg)^3 
\delta(v-\xi)\delta(1-\frac{ z}{x}) O^{i,a}\, , \nn 
\eea
where again $a$ is not summed over and the diagrams related to $C$ by symmetry are obtained by making the  substitutions in Eq.~(\ref{subs2}).
The result of summing diagram $C$ and its reflections is
\bea
M_C^{i,a}+M_{\bar C}^{i,a}+M_{\hat C}^{i,a}+M_{\bar{\hat C}}^{i,a}&=& -\frac{\alpha_s}{2\pi} \frac{\beta^{(a)}}{\epsilon}
\bigg\{\bigg[ \frac{4}{\eta}+\ln\bigg(\frac{z^4 \nu}{u\bar u v \bar v P^4}\bigg)\bigg]\delta(\lambda-u)\delta(\xi-v) \\
&&\hspace{-2ex}-\bigg[\frac{u}{\lambda}\frac{\theta(\lambda-u)}{(\lambda-u)_+} + \frac{\bar u}{\bar \lambda}\frac{\theta(u- \lambda)}{(\bar \lambda-\bar u)_+}\bigg]\delta(\xi-v) \nn  \\
&&\hspace{-2ex}
- \bigg[\frac{v}{\xi}\frac{\theta(\xi-v)}{(\xi-v)_+}
+\frac{\bar v}{\bar \xi}\frac{\theta(v-\xi)}{(\bar \xi-\bar v)_+}\bigg]  \delta(\lambda-u)\bigg\} 
 \bigg(\frac{z}{2P}\bigg)^3 
 \delta(1-z/x) O^{i,a}  \, . \nn
\eea

The virtual diagrams and their reflections do not lead to any mixing between singlet and octet operators or between Dirac structures 1 and 2. 
The $1/\eta$ poles and the corresponding logarithms of $\nu$ cancel between diagrams $B$ and $C$ and their reflections when the operator is a color-singlet, but not when it is a color-octet. In the color-octet case the rapidity divergences cancel against rapidity divergences in the real emission graphs which we evaluate next.
 
Real radiation comes from diagrams $D-G$ and their reflections.
The color factors for these diagrams are given by:
\[
\beta_{ab}=\left(
\begin{array}{cc}
 \beta_{11}  & \beta_{18}     \\
  \beta_{81}& \beta_{88}    \\  
\end{array}
\right)
=\left(
\begin{array}{cc}
 0  &  \frac{C_F}{2N_c}     \\
1 & \frac{N_c^2-2}{2N_c}     \\  
\end{array}
\right)
,\qquad
\bar \beta_{ab}=\left(
\begin{array}{cc}
 0  &    \frac{C_F}{2N_c}   \\
 1  & -\frac{1}{N_c}     \\  
\end{array}
\right) \, .
\]
The first index, $a=1$ or $8$,  refers to the color state of the initial and final  state quarks in the diagram and the second index, $b$, refers to the color-structure of the operator. The real radiation diagrams shown in Fig.~\ref{oneloopfd} evaluate to: 
\bea
M_D^{i,a}&=& \frac{\alpha_s}{2\pi} \left( \frac{z}{2P}\right)^3 \frac{1}{\epsilon_{UV}} \beta_{ab} \frac{x^2}{z^2}\frac{1-z/x}{\lambda \xi} \delta\left(\bar v-\frac{z}{x}\bar \xi\right) \delta\left(\bar u -\frac{z}{x}\bar \lambda\right)\frac{\left(O^{1,b} + O^{2,b}\right)}2,
\\\nonumber\\
M_E^{i,a}&=& (-1)^i\frac{\alpha_s}{2\pi} \left( \frac{z}{2P}\right)^3 \frac{1}{\epsilon_{UV}}\bar  \beta_{ab} \frac{x^2}{z^2}\frac{1-z/x}{\lambda \bar \xi} \delta\left(v-\frac{z}{x} \xi \right) \delta\left(\bar u -\frac{z}{x}\bar \lambda\right)\frac{\left(O^{1,b} - O^{2,b}\right)}2,
\nonumber\\\nonumber\\
M_F^{i,a} &=& \frac{\alpha_s}{2\pi} \left( \frac{z}{2P}\right)^3 \frac{1}{\epsilon_{UV}} \beta_{ab}\bigg\{ -\bigg[ \frac{1}{\eta}+\ln\bigg(\frac{z \nu}{uP}\bigg)\bigg]\delta(1-z/x)
+\frac{u x}{\lambda z}\frac{\theta(1-z/x)}{ (1-z/x)_+} \bigg\}\nn\\
&& \hspace{25ex}\times \delta\left(\bar v-\frac{z}{x}\bar \xi\right) \delta\left(\bar u -\frac{z}{x}\bar \lambda\right)O^{i,b},
\nonumber\\\nonumber\\
M_G^{i,a}&=&- \frac{\alpha_s}{2\pi} \left( \frac{z}{2P}\right)^3 \frac{1}{\epsilon_{UV}}\bar  \beta_{ab} \bigg\{ -\bigg[ \frac{1}{\eta}+\ln\bigg(\frac{z\nu}{u P}\bigg)\bigg]\delta(1-z/x)
+\frac{u x}{\lambda z}\frac{\theta(1-z/x)}{ (1-z/x)_+}\bigg\}\nn\\
&& \hspace{25ex}\times \delta\left(v-\frac{z}{x}\xi\right) \delta\left(\bar u -\frac{z}{x}\bar \lambda\right)O^{i,b}
\, , \nn
\eea
where $b$ is not summed over.  The reflections of $D-G$ are given by
the substitutions in Eq.~(\ref{subs2}). The real diagrams mix color-singlet and color-octet operators, and diagrams $D$ and $E$ also mix the Dirac structures.

%\section{Renormalization}
The DPFF is renormalized multiplicatively,
\beq
D^{Q\bar Q(\textrm{\small bare})}_{i,a}(u,v,z)= \int du^\prime dv^\prime \frac{dz^\prime}{z^\prime} Z_{ia,jb}(u,u^\prime,v,v^\prime,z/z^\prime,\mu)D^{Q\bar Q}_{j,b}(u^\prime,v^\prime,z^\prime,\mu)\, ,
\eeq
and thus obeys the renormalization group equation
\beq
\mu \frac{d}{d\mu} D^{Q\bar Q}_{i,a}(u^{\prime \prime},v^{\prime \prime},z^{\prime \prime},\mu)= - \int 
du^\prime dv^\prime \frac{dz^\prime}{z^\prime}\gamma_{ia,jb} (u^{\prime \prime},u^\prime,v^{\prime \prime},v^\prime,z^{\prime \prime}/z^\prime,\mu)D^{Q\bar Q}_{j,b}(u^{\prime},v^{\prime},z^{\prime},\mu)\,,
\eeq
where the  anomalous dimension is given by
\bea
\gamma_{ia,jb}(u,u^\prime,v,v^\prime,z/z^\prime,\mu)&=& \\
&& \hspace{-10ex}\int du^{\prime \prime}dv^{\prime \prime}\frac{dz^{\prime \prime}}{z^{\prime \prime}} Z_{ia,kc}^{-1}(u,u^{\prime \prime},v,v^{\prime \prime},z/z^{\prime \prime},\mu)\mu\frac{d}{d\mu}
Z_{kc,jb}(u^{\prime \prime},u^\prime,v^{\prime \prime},v^\prime,z^{\prime \prime}/z^\prime,\mu) \, .\nn
\eea
The indices $i$ and $j$ label the Dirac structure and the indices $a$ and $b$ refer to color $(a,b=1\,\textrm{or}\,8)$.
The tree-level matrix element of the DPFF using partonic states with momenta labelled by $(x,\lambda,\xi)$ is given by
\bea
D^{Q\bar Q}_{j,b}(u,v,z)=\left( \frac{z}{2P}\right)^3 \delta(1-z/x)\delta(\lambda-u)\delta(\xi-v)O^{j,b} \,.
\eea
Finally, we need to include
the wave function renormalization,
\beq
Z_\xi=1- \frac{\alpha_s C_F}{4 \pi \epsilon},
\eeq
after which we find that the anomalous dimensions are: 
\bea
\gamma_{11,11}  &=&-\frac{\alpha_s C_F}{\pi}   \delta(1-z/z^\prime) \bigg( 3 \delta(u-u^\prime)\delta(v-v^\prime) \\
&&\qquad+ \delta(v-v^\prime)\left\{ \theta(u^\prime-u) \frac{u}{u^\prime}\left[ \frac{1}{(u^\prime-u)_+} + 1\right] + \theta(u-u^\prime) \frac{\bar u}{\bar u^\prime} \left[\frac{1}{(u-u^\prime)_+} + 1\right]\right\}\nn \\
&&\qquad+ \delta(u-u^\prime)\left\{\theta(v^\prime-v) \frac{v}{v^\prime} \left[ \frac{1}{(v^\prime-v)_+}+ 1\right] + \theta(v-v^\prime) \frac{\bar v}{\bar v^\prime} \left[\frac{1}{(v-v^\prime)_+}+1 \right]\right\}\bigg), \nn\\
&&\nn\\
\gamma_{21,21} &=& \gamma_{11,11}\, , \\
\gamma_{18,11} &=& - \frac{\alpha_s}{\pi} \theta(1-z/z^\prime) \left(\frac{z}{z'}\right)^2 \left\{\left[  \frac{u v' +v u'}{ (1-z/z')_+} + \frac{1-z/z'}{2z/z'}\right] \frac1{u'v'} \delta(\bar v-\frac{z}{z'}\bar v') \delta(\bar u -\frac{z}{z'}\bar u')\right.\nn  \\
&& \phantom{- \frac{\alpha_s}{\pi}  \frac{z'}{z}} +\left[    \frac{\bar u\bar v' +\bar v\bar u' }{(1-z/z')_+}     + \frac{1-z/z'}{2z/z'}\right]  \frac{1}{\bar u'\bar v'} \delta(v-\frac{z}{z'}v') \delta(u -\frac{z }{z'}u') \nn\\
&& \phantom{- \frac{\alpha_s}{\pi}  \frac{z'}{z}}  -\left[   \frac{u\bar v' +\bar v u'   }{(1-z/z')_+}  + \frac{1-z/z'}{2z/z'} \right] \frac1{u'\bar v'} \delta(v-\frac{z}{z'} v' ) \delta(\bar u -\frac{z}{z'}\bar u')\nn\\
&&\left.\phantom{- \frac{\alpha_s}{\pi}  \frac{z'}{z}} - \left[  \frac{\bar u v' + v \bar u' }{(1-z/z')_+} + \frac{1-z/z'}{2z/z'} \right] \frac1{\bar u' v'} \delta(\bar v-\frac{z  }{z'}\bar v') \delta(u -\frac{z }{z'}u')\right\},
\\
&&\nn\\
\gamma_{18,21}&=& - \frac{\alpha_s}{\pi} \theta(1-z/z^\prime) \left(\frac{z}{z'}\right)^2 \left\{\left[  \frac{u v' +v u'}{ (1-z/z')_+} + \frac{1-z/z'}{2z/z'}\right] \frac1{u'v'} \delta(\bar v-\frac{z}{z'}\bar v') \delta(\bar u -\frac{z}{z'}\bar u')\right.\nn  \\
&& \phantom{- \frac{\alpha_s}{\pi}  \frac{z'}{z}} +\left[    \frac{\bar u\bar v' +\bar v\bar u' }{(1-z/z')_+}     + \frac{1-z/z'}{2z/z'}\right]  \frac{1}{\bar u'\bar v'} \delta(v-\frac{z}{z'}v') \delta(u -\frac{z }{z'}u') \nn\\
&& \phantom{- \frac{\alpha_s}{\pi}  \frac{z'}{z}}  -\left[   \frac{u\bar v' +\bar v u'   }{(1-z/z')_+}  - \frac{1-z/z'}{2z/z'} \right] \frac1{u'\bar v'} \delta(v-\frac{z}{z'} v' ) \delta(\bar u -\frac{z}{z'}\bar u')\nn\\
&&\left.\phantom{- \frac{\alpha_s}{\pi}  \frac{z'}{z}} - \left[  \frac{\bar u v' + v \bar u' }{(1-z/z')_+} - \frac{1-z/z'}{2z/z'} \right] \frac1{\bar u' v'} \delta(\bar v-\frac{z  }{z'}\bar v') \delta(u -\frac{z }{z'}u')\right\},
\\
\gamma_{28,11}&=& \gamma_{18,21},\\
\gamma_{28,21} &=& \gamma_{18,11}.
\eea
In addition, 
\beq
\gamma_{i1,j8} = \frac{C_F}{2N_c} \gamma_{i8,j1} \, .
\eeq
Furthermore, 
\bea
\gamma_{18,18} &=&  -3 \frac{\alpha_sC_F}{\pi} \delta(u-u^\prime)\delta(v-v^\prime) \delta(1-z/z^\prime)\\
&&\hspace{-1ex}+\frac{\alpha}{\pi}\frac{1}{2N_c}\delta(v-v^\prime)\delta(1-\frac{ z}{z^\prime})\left[ \theta(u^\prime-u) \frac{u}{u^\prime}\left( \frac{1}{(u^\prime-u)_+}+1\right) + (u\leftrightarrow \bar u, u'\leftrightarrow \bar u') \right]\nn \\
&&\hspace{-1ex}+ \frac{\alpha_s}{\pi}\frac{1}{2N_c}\delta(u-u^\prime)\delta(1-\frac{ z}{z^\prime}) \left[\theta(v^\prime-v) \frac{v}{v^\prime}\left(\frac{1}{(v^\prime-v)_+}+1\right)  + (v \leftrightarrow \bar v, v' \leftrightarrow \bar v') \right]\nn\\
&&\hspace{-1ex} - \frac{\alpha_s}{\pi} \left(\frac{z}{z'}\right)^2 \left\{\frac{N_c^2 - 2}{2 N_c} \left[  \frac{ u v' +v u'}{ (1-z/z')_+} + \frac{1-z/z'}{2z/z'}  \right] \frac1{u'v'} \delta(\bar v-\frac{z}{z'}\bar v') \delta(\bar u -\frac{z}{z'}\bar u')\right.  \nn\\
&&\hspace{-1ex} \phantom{-\frac{\alpha_s}{\pi} \frac{z'}{z} }  +\frac{N_c^2 - 2}{2 N_c}\left[   \frac{ \bar u\bar v' +\bar v\bar u'}{(1-z/z')_+}     + \frac{1-z/z'}{2z/z'}  \right]  \frac{1}{\bar u'\bar v'} \delta(v-\frac{z}{z'}v') \delta(u -\frac{z }{z'}u') \nn\\
&&\hspace{-1ex} \phantom{-\frac{\alpha_s}{\pi} \frac{z'}{z} }  +\frac1{N_c}\left[    \frac{u\bar v'  +\bar v u'  }{(1-z/z')_+}  + \frac{1-z/z'}{2z/z'}   \right] \frac1{u'\bar v'} \delta(v-\frac{z}{z'} \xi ) \delta(\bar u -\frac{z}{z'}\bar u')\nn\\
&&\hspace{-1ex}\left. \phantom{-\frac{\alpha_s}{\pi} \frac{z'}{z} } +\frac1{N_c} \left[ \frac{\bar u v' + v \bar u' }{(1-z/z')_+} + \frac{1-z/z'}{2z/z'}\right] \frac1{\bar u' v'} \delta(\bar v-\frac{z  }{z'}\bar v') \delta(u -\frac{z }{z'}u')\right\} \theta(1-z/z^\prime) , \nn
\eea
\bea
\gamma_{18,28} &=&  - \frac{\alpha_s}{\pi} \left(\frac{z}{z'}\right)^2 \left\{\frac{N_c^2 - 2}{2 N_c} \left[  \frac{ u v' +v u'}{ (1-z/z')_+} + \frac{1-z/z'}{2z/z'}  \right] \frac1{u'v'} \delta(\bar v-\frac{z}{z'}\bar v') \delta(\bar u -\frac{z}{z'}\bar u')\right.  \\
&&\hspace{-1ex} \phantom{-\frac{\alpha_s}{\pi} \frac{z'}{z} }  +\frac{N_c^2 - 2}{2 N_c}\left[   \frac{ \bar u\bar v' +\bar v\bar u'}{(1-z/z')_+}     + \frac{1-z/z'}{2z/z'}  \right]  \frac{1}{\bar u'\bar v'} \delta(v-\frac{z}{z'}v') \delta(u -\frac{z }{z'}u') \nn\\
&&\hspace{-1ex} \phantom{-\frac{\alpha_s}{\pi} \frac{z'}{z} }  +\frac1{N_c}\left[    \frac{u\bar v'  +\bar v u'  }{(1-z/z')_+}  - \frac{1-z/z'}{2z/z'}   \right] \frac1{u'\bar v'} \delta(v-\frac{z}{z'} \xi ) \delta(\bar u -\frac{z}{z'}\bar u')\nn\\
&&\hspace{-1ex}\left. \phantom{-\frac{\alpha_s}{\pi} \frac{z'}{z} } +\frac1{N_c} \left[ \frac{\bar u v' + v \bar u' }{(1-z/z')_+} - \frac{1-z/z'}{2z/z'}\right] \frac1{\bar u' v'} \delta(\bar v-\frac{z  }{z'}\bar v') \delta(u -\frac{z }{z'}u')\right\} \theta(1-z/z^\prime) , \nn
\\
\gamma_{28,18} &=& \gamma_{18,28}\\
\gamma_{28,28} &=& \gamma_{18,18}.
\eea

\begin{figure}
\begin{center}
\includegraphics[width=2.5in]{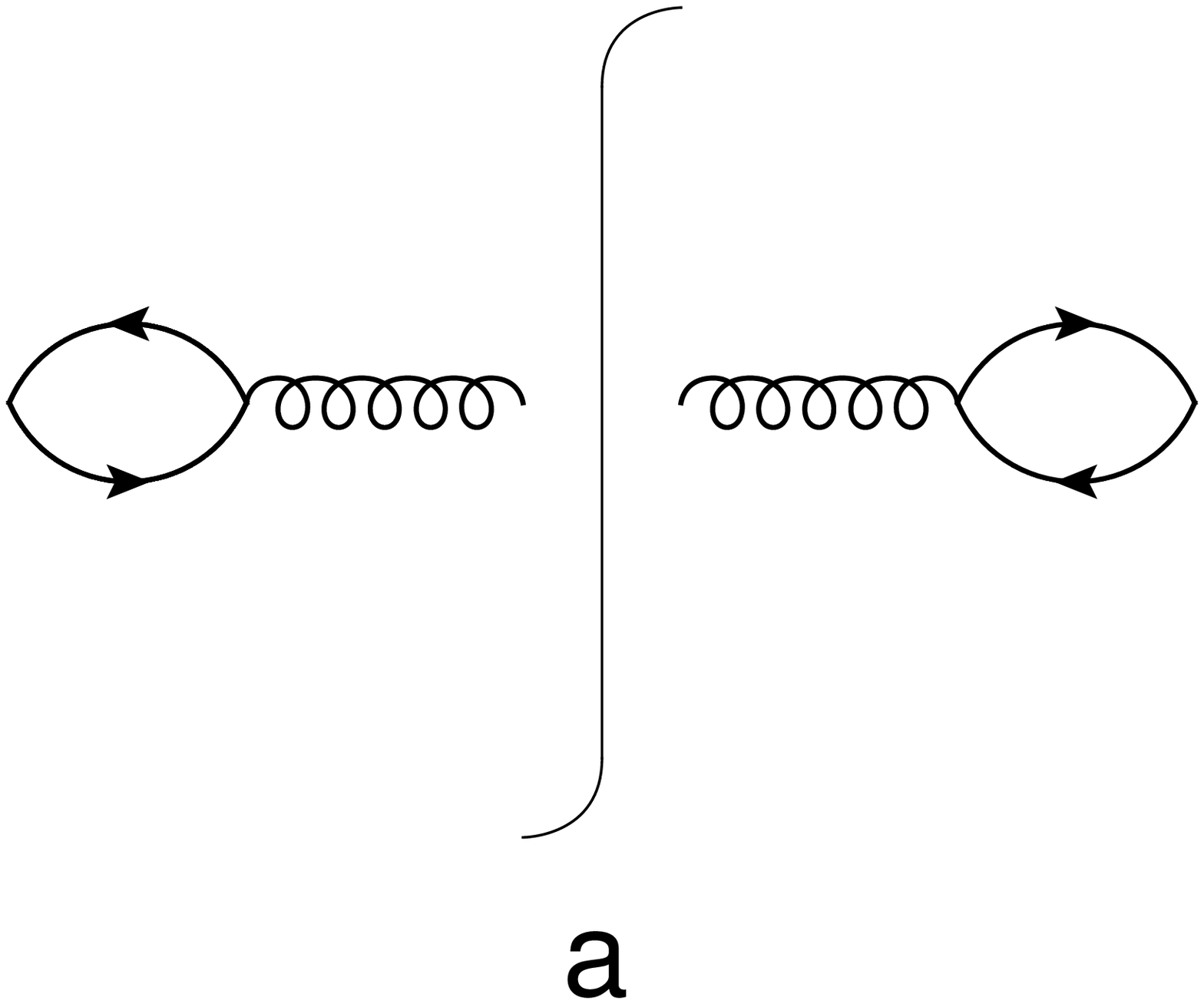}\qquad\includegraphics[width=2.5in]{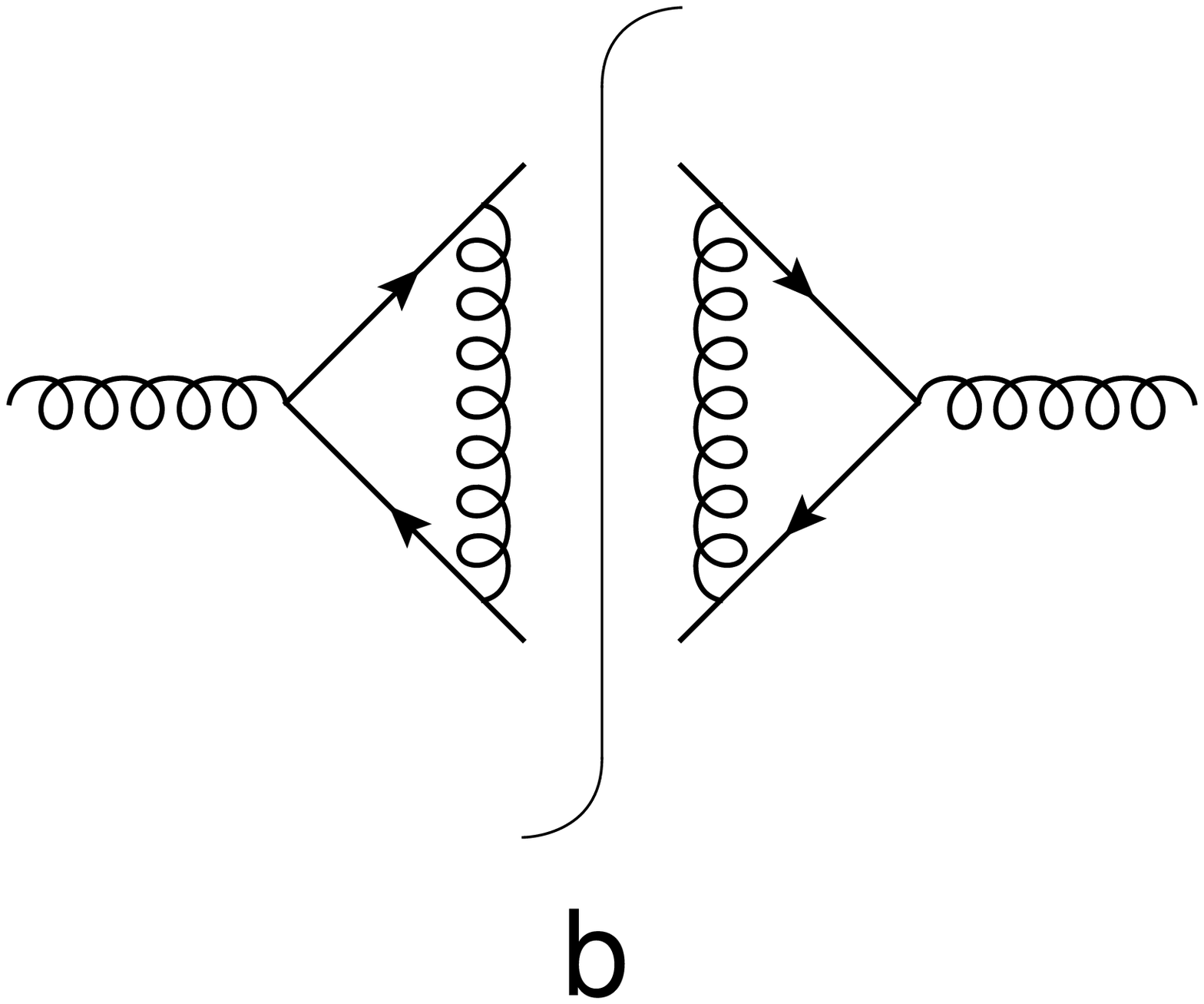}\caption{Possible mixings between single and double parton fragmentation functions.   }
\label{submixing}
\end{center}
\end{figure}

In addition to the above mixing matrix we must also consider the mixing with power suppressed
single parton fragmentation functions.  As emphasized in the introduction, while the direct
contribution of such operators will not receive any $1/v$ enhancement,  in principle
they can still affect the runnings of the DPFFs. For instance we may consider the mixing
into the power suppressed single gluon fragmentation function.  Such mixing can in general
arise from diagrams such as those in Fig.~\ref{submixing}a. However, to mix from the single gluon
 to the double quark fragmentation function would require diagrams such as those in 
 Fig.~\ref{submixing}b, which is sub-leading in $\alpha(p_\perp)$.  Thus at lowest order the anomalous dimension
 matrix will contain only one off-diagonal element that will not generate any logs proportional
 to the DPFF. Then, given that the single gluon power
 suppressed fragmentation function is not enhanced we may ignore any mixing between
 double and power suppressed single fragmentation functions.

%\section{Conclusions}

To summarize, we have extended the calculation of the anomalous dimensions of the DPFF in Ref.~\cite{Fleming:2012wy} to include 
all possible Dirac structures. This gives the complete set of anomalous dimensions needed at this order in $p_\perp/m_Q$. The renormalization group equations presented here and in Ref.~\cite{Fleming:2012wy} can be evolved from the scale $p_\perp$ to $m_Q$ to resum logarithms of $p_\perp/m_Q$ in double parton fragmentation contributions to quarkonium production. When this is completed, it will be interesting to see if the resummed cross sections can improve our understanding of quarkonium production.

%%%%%%%%%%%%%%%%%%%%%%%%%%%%%%%%%%%%%%%%
% APPENDIX
%%%%%%%%%%%%%%%%%%%%%%%%%%%%%%%%%%%%%%%%
%\appendix

\acknowledgments 

SF  was supported in part by the Director, Office of Science, Office of Nuclear Physics, of the U.S. Department of Energy under grant numbers DE-FG02-06ER41449 and DE-FG02-04ER41338. SF  also acknowledges support from the DFG cluster of excellence ``Origin and structure of the
universe''. AKL was supported in part by the National Science Foundation under Grant No. PHY-1212635. TM was supported in part by the Director, Office of Science, Office of Nuclear Physics, of the U.S. Department of Energy under grant numbers DE-FG02-05ER41368.  IZR  is supported by 
DOE DE-FG02-04ER41338 and FG02-06ER41449.

%%%%%%%%%%%%%%%%%%%%%%%%%%%%%%%%%%%%%%%%%
%Bibliography

\bibliography{mixing}

\end{document}